\documentclass{iopart}
\usepackage{amsmath,amssymb}
\usepackage[T1]{fontenc}
\usepackage{graphicx}
\usepackage{hyperref}
\usepackage[dvipsnames]{xcolor}
\usepackage{comment}

\begin{document}

\title{GWnext 2024: Meeting Summary}
\author{Alejandro Torres-Orjuela, Ver{\'o}nica V{\'a}zquez-Aceves, Rui Xu, Jin-Hong Chen, Andrea Derdzinski, Matthias U. Kruckow, Stefano Rinaldi, Lorenzo Speri, Ziming Wang, Garvin Yim, Xue-Ting Zhang, Qian Hu, Miaoxin Liu, Xiangyu Lyu, Zheng Wu, Cong Zhou, Manuel Arca Sedda, Yan-Chen Bi, Hong-Yu Chen, Xian Chen, Jiageng Jiao, Yu-Mei Wu}
\ead{atorreso@hku.hk}

\begin{indented}
\item[]\today
\end{indented}

\begin{abstract}
GWnext 2024 was a meeting held in the Kavli Institute for Astronomy and Astrophysics at Peking University in March $4^\text{th} - 8^\text{th}$, 2024. In the meeting researchers at different career stages -- with a particular focus on early career scientists -- working on the different aspects of gravitational wave (GW) astronomy gathered to discuss the current status as well as prospects of the field. The meeting was divided into three core sessions: Astrophysics, GW Theory, and Detection. Each session consisted of introductory talks and extended discussion sessions. Moreover, there was a poster session where students could present their results. In this paper, we summarize the results presented during the meeting and present the most important outcomes.
The webpage of the meeting can be found at:
\begin{itemize}
    \item \href{https://aletorreso.wixsite.com/atogw/gwnext2024}{https://aletorreso.wixsite.com/atogw/gwnext2024}
    \item \href{https://kiaa.pku.edu.cn/gwave/Home.htm}{https://kiaa.pku.edu.cn/gwave/Home.htm}
\end{itemize}
\end{abstract}

\section{Introduction}\label{sec:int}

\href{https://aletorreso.wixsite.com/atogw/gwnext2024}{GWnext 2024} was the second edition in a series of meetings on gravitational wave (GW) astronomy. The meeting was held at the \href{http://kiaa.pku.edu.cn/}{Kavli Institute for Astronomy and Astrophysics} at Peking University in March $4^\text{th} - 8^\text{th}$, 2024. The general goal of GWnext is to bring people at different career stages -- with a particular focus on early career scientists -- working on the different aspects of GW astronomy together to discuss the current status of the field and the prospects in the coming years. The goal is to have a vivid exchange between all participants to foster collaborations, in particular between people from different fields, as well as to stimulate the discussion of new ideas. Therefore, the number of talks was reduced to a small number, and discussion sessions made up a significant part of the meeting.

The meeting was divided into three core sessions on Astrophysics, GW Theory, and Detection. Each session lasted one day starting with three overview talks to give a general introduction to all participants followed by extended discussion sessions. Before the meeting, the participants were asked to provide at least two questions that were used as a thread during the discussion sessions. These sessions were moderated by chairs but encouraged a free and spontaneous interaction between the participants. They lasted between two to three hours and were only closed when the audience signaled satisfaction with the discussion of all questions. Moreover, a poster session where students could present their work was organized. The poster session lasted for around two hours and all participants of the meeting were asked to vote for the best posters to encourage interaction with the students presenting. The presenters of the three posters that got the most votes were awarded prizes on the last day of the meeting.

In this paper, we give an overview of the meeting and summarize the most important outcomes. In Section~\ref{sec:talk}, the overview talks are summarized while summaries of the posters are presented in Section~\ref{sec:post}. The most important outcomes of the discussion sessions are presented in Section~\ref{sec:dis}. In Section~\ref{sec:sum}, we make conclusions about the meeting and discuss briefly expectations for the future. The full list of questions contributed by the participants and the affiliations of the authors can be found in the Appendix.

\section{Overview talks}\label{sec:talk}

\subsection{Astrophysics}

\paragraph{\textbf{Andrea Derdzinski} -- Science potential of gas-rich galactic nuclei for future GW experiments} \mbox{}

Active galactic nuclei (AGN) are promising environments for GW sources across the detectable spectrum, especially at sub-Hz frequencies. Gas accretion onto supermassive black holes (SMBHs) produces bright emissions observed in AGN. The influence of the gaseous disc in these environments can lead to the formation of SMBH binaries, stellar-origin black hole binaries (BHBs), and the inspiral of stellar-origin black holes (BHs) with the central SMBH.  The ability for relevant sources to be both loud (especially for SMBH binaries) and drive bright emissions (especially in near-Eddington AGN systems) presents a promising case for multimessenger discoveries. With the use of simplified disc structure models~\cite{shakura_1973}, we can gain intuition on the rates and gas-imparted characteristics of these sources, which becomes particularly exciting for future space-based GW detections. 

\textit{SMBH binaries:} SMBH binaries may form as a consequence of galaxy mergers, after a series of dynamical interactions that drive constituent SMBHs into the center of the remnant galaxy. Major galaxy mergers are additionally expected to funnel gas into the nucleus, if present, enabling the interaction of traveling SMBHs with fresh accretion flows. This interplay is a critical stage in the formation of a SMBH binary in addition to stellar hardening (see Ref.~\cite{lisa_2022}). Naturally, we expect that some fraction of observed AGN may harbor SMBH binary systems. This understanding motivates electromagnetic (EM) campaigns to search for signatures of SMBH binary-disc interaction, which has led to a series of binary candidates via radio very-long-baseline interferometry (VLBI) as well as photometric or spectroscopic variability (see the review~\cite{d'orazio_2023}). 

Recent analysis by pulsar timing arrays (PTAs) provides evidence for a cosmological population of the most massive SMBHs via constraints on the nHz GW background~\cite{agazie_2023}. In the future, upcoming space-based detectors will enable detections of single SMBH binary systems with impressive signal-to-noise ratio (SNR) ($O(100)$), allowing for precise measurements of the SMBH masses, spins, and inspiral orbital properties~\cite{lisa_2024}. The interaction with gas for these sources is of particular importance for nHz to mHz sources, as it has implications for the distribution of binary mass ratios~\cite{duffel_2020}, SMBH spins~\cite{steinle_2023}, and eccentricity at sub-parsec stages~\cite{siwek_2023,d'orazio_2021}. Gas-induced deviations may also be measurable in the waveforms of mHz sources, providing a unique opportunity to probe disc interaction with GWs~\cite{garg_2022}. For recent analysis with the Laser Interferometer Space Antenna (LISA) detections, we refer the reader to Figs. 3 and 4 of Ref.~\cite{garg_2024}, which highlight the detectable parameter space for both eccentricity and gas deviations, calculated via the Fisher formalism and confirmed with Bayesian inference.

\textit{Stellar-origin BH mergers:} Aside from post-merger galactic nuclei, active single-SMBH systems are also interesting for GW source production from mHz to Hz frequencies, especially in systems that contain nuclear stellar clusters. In these dense stellar environments, the interaction of stars or their remnants with the accretion disc will impact their orbital evolution. If sufficient orbit intersections occur through sufficiently dense gas, the stellar orbit inclination will reduce to the plane of the disc~\cite{syer_1991,fabj_2020}. Stars can also form within the disc itself via gravitational instability, leading to an additional source of embedded BHs~\cite{levin_2007,derdzinski_2023}.  When taking predicted stellar-origin BH merger rates from semi-analytical models~\cite{tagawa_2020,groebner_2020} divided by the observationally inferred rate~\cite{ligo_2023}, we find that only a fraction of currently detected events could be produced in AGN. Whether embedded BHs find each other within the disc depends on sensitive migration physics and disc structure~\cite{bellovary_2016}, and the above works neglect this complexity which can lead to an even larger uncertainty. For a glimpse into the hurdles imparted by feedback or disc turbulence, we refer to Fig. 6 in Ref.~\cite{grishin_2023}, as well as the simulations presented in Ref.~\cite{wu_2024}, respectively. 

Rates aside, sources in this regime carry interesting consequences from disc interaction. Counterintuitively, the sources can become highly eccentric during gas-driven binary capture~\cite{fabj_2024}. More sensitive future detectors may be able to place limits on eccentricity~\cite{saini_2024}, although a more promising case lies in lower frequency detections of binaries at earlier inspiral stages. Decihertz detectors have a strength in this regard, for example, the Lunar Gravitational Wave Antenna~\cite{harms_2021}, the Laser Interferometer Lunar Antenna (LILA)~\cite{jani_2024}, or atom interferometer obsevatories~\cite{abend_2023}.

\textit{E/IMRIs:} The mechanisms described above for stellar-origin BHBs are also relevant for the production of extreme mass ratio inspirals (EMRIs) or intermediate mass ratio inspirals (IMRIs), detectable with space-based missions probing the mHz band. As BHs travel towards the central SMBH, they can grow via gas accretion and potential hierarchical mergers. The rate predictions here suffer from the same uncertainty as stellar-origin BH merger estimates, although the BHs, in this case, avoid the primary hurdle of having to find each other to form a tight binary (see e.g. Ref.~\cite{rowan_2024}). As gas accretes towards the SMBH, it naturally may bring embedded BHs with it. Given the ability of gas to facilitate BH interaction that would otherwise be limited by loss-cone scattering, AGN are anticipated to experience a boost in the production rate of E/IMRIs~\cite{pan_2021b,derdzinski_2023}. The formation of gas-driven E/IMRI events is ultimately dependent on the details of the BH population residing in AGN discs, and thus the detection rates of stellar-origin BH mergers must be connected with subsequent E/IMRI rates. With the advent of increasingly sensitive ground-based detectors, we can begin to place limits on formation pathways for the population of merging stellar-origin BHs, although more work remains to be done to understand the efficiency of BH capture versus in-situ formation, subsequent BH interaction, and dependence on the type of AGN system. 

The gas-driven formation scenario is expected to primarily produce low eccentricity E/IMRIs (although exceptions are possible, see e.g. Ref.~\cite{secunda_2021}). These sources are also expected to experience environmental perturbations from the surrounding gas, which are dependent on the disc structure and response to embedded perturbers~\cite{yunes_2011,derdzisnki_2019,derdzinski_2021}. These perturbations may interfere with precise parameter estimation~\cite{barausse_2014,zwick_2023}, yet also provide valuable insight into the details of disc-BH interaction which are otherwise difficult to probe electromagnetically.

\paragraph{\textbf{Matthias U. Kruckow} -- Imprints of stellar binary evolution on the population observed by GWs} \mbox{}

The so far observed GW sources are believed to mainly result from binary stellar evolution. Because of the strong dependence of GW radiation on the distance between two orbiting bodies, non-interacting stars will not be able to merge due to GW radiation. Hence, all the observed events have some interacting binaries as progenitors. There are different kinds of interaction starting from tidal effects over mass transfer to impact on the companion of ejected material, e.g. in a supernova. A very uncertain but important interaction to produce short-period compact-object binaries is the so-called common envelope evolution. In this phase, the envelope material of the donating star(s) surrounds both components and leads to a strong orbital shrinkage while ejecting the envelope material.

Each kind of interaction can cause some imprints on observables derived from GW signals. While the kind of mass transfer and therefore the amount of accreted material will alter the spin magnitude in the merging event, the kick of the supernova events will impact the orientation of the spin alignments. Both will enter in the observable effective spin. The observable mass ratio will be altered significantly only by the mass transfer. Hence, relations between the mass ratio and the effective spin will most likely originate from mass transfer, while effects on the spin alone may tell us more about the kicks during the compact object formation or tidal effects (see e.g. Figs. 2 and 5 in Ref.~\cite{bavera_2021}).

The rate of different classes of compact object mergers have relics of the effectiveness of binary interactions imprinted (cf. Figs. 3.47, and B.46 to B.51 in Ref.~\cite{kruckow_2018}). This especially holds for ratios of rates to overcome normalization uncertainties introduced by e.g. the star formation rate and the metallicity evolution over cosmic time. Other physical processes involved in the supernova during compact object formation can lead to mass gaps, like the one expected between neutron stars and BHs or the BH mass gap expected from pair instability (see e.g. Fig. 15 in Ref.~\cite{briel_2023}).

\paragraph{\textbf{Jin-Hong Chen} -- EM and gravitational radiation in tidal disruption events} \mbox{}

Occasionally, a star in the nucleus of a galaxy can be perturbed into an orbit on which it comes too close to the central massive BH and it gets destroyed by its tidal force; this is a so-called tidal disruption event (TDE). The stellar debris is then pulled into an elongated structure that collides in self-intersection due to relativistic effects, which results in the circularization of the orbit. The self-intersection of the debris stream might be crucial for the formation of an accretion disk. Usually, the accretion process is super-Eddington at first and turns into sub-Eddington accretion as the accretion rate decreases. A bright flare produced by the collision of the debris stream and the accretion process can reveal a massive BH at the center of a galaxy or a star cluster. 

So far, X-ray telescopes and optical observatories have detected over 100 TDE candidates (see the reviews in Refs.~\cite{vanvelzen_2020,saxton_2020}). Recent advancements have expanded TDE research to include observations in different wavelengths, such as delayed radio flares~\cite{alexander_2016}, dust echo mid-infrared emissions~\cite{masterson_2024}, and optical polarization detection~\cite{liodakis_2023}. In 2021, the IceCube team reported the detection of neutrinos from a TDE candidate~\cite{stein_2021}, ushering TDEs into a multi-messenger era. This development raises the question of whether we can also detect GW radiation from TDEs in the future.

\begin{figure}\centering
\includegraphics[width=0.58\textwidth]{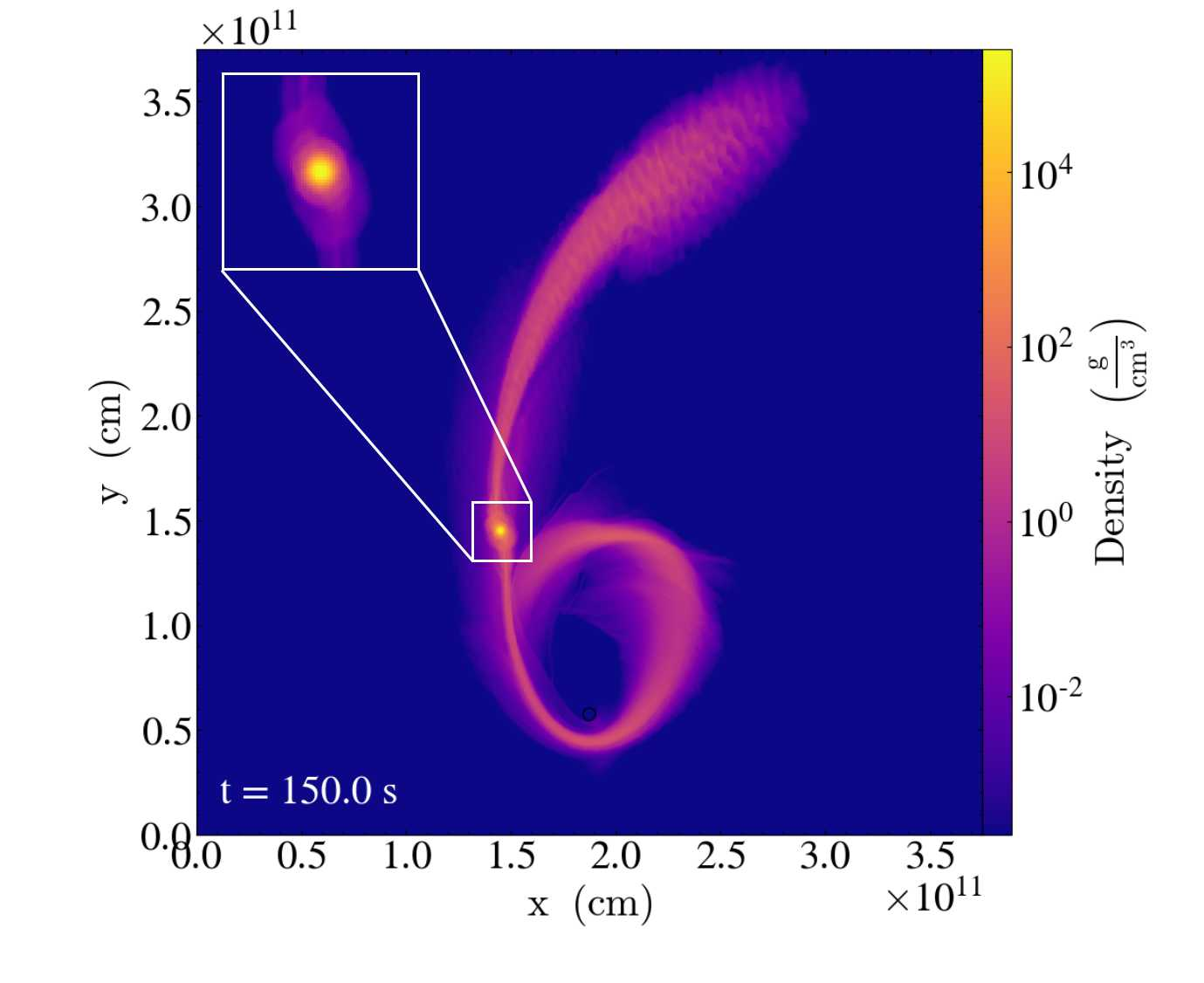}
    \caption{
        Projection of the gas density on the orbital plane for the tidal stripping of a WD by an IMBH in an eccentric orbit. The first returning debris has passed through the pericenter and formed a ring-like structure. The WD which is shown in the inset diagram is returning to the pericenter. The simulation is conducted using the FLASH hydrodynamics code. The time since the beginning of the simulation is shown in the lower left corner. The IMBH is indicated by the black circle. Figure from Ref.~\cite{chen_2023}.
    }\label{fig:wd}
\end{figure}

\begin{figure}\centering
\includegraphics[width=0.58\textwidth]{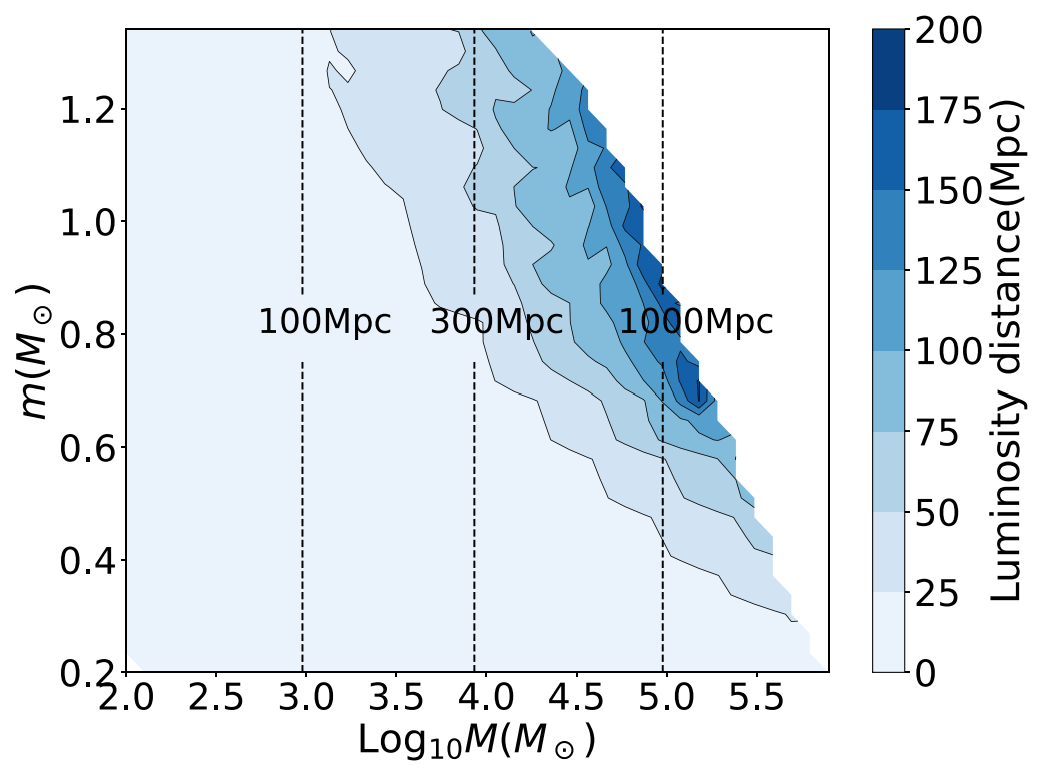}
    \caption{
        The blue shadowed region represents the GW horizon distance for a WD-TDE system detected by TianQin. The dark dashed lines are the maximum detection distance for the EM emission from the horizon distance for WD-TDEs using Einstein Probe. The luminosity of the EM emission is capped at the Eddington luminosity of the central BH. Figure from Ref.~\cite{ye_2024}.
    }\label{fig:tdedist}
\end{figure}

Recent research has explored the GW detectability of TDEs~\cite{toscani_2022,pfister_2022}. While detecting GW signals from on-off TDEs is challenging, there is promise in detecting GW signals from repeating TDEs using next-generation GW detectors. Of particular interest are repeating partial TDEs of a white dwarf by an IMBH, which could also be the origin of quasi-periodic eruptions (QPEs)~\cite{macleod_2014,king_2020,miniutti_2023,chen_2023}. See Fig.~\ref{fig:wd} for the tidal stripping of a white dwarf (WD) by an IMBH. These events may emit bright EM bursts and GW emissions within a few years. The detectable range for these emissions could extend up to 1000 Mpc for EM signals and 200 Mpc for GW signals~\cite{ye_2024} as shown in Fig.~\ref{fig:tdedist}. With the advent of numerous new and highly sensitive telescopes, as well as advancements in GW detectors, it is anticipated that they will not only detect a greater number of TDE candidates, but also usher in a new era of multi-messenger astronomy in the study of TDEs.

\subsection{GW theory}

\paragraph{\textbf{Garvin Yim} -- Continuous GWs from isolated neutron stars} \mbox{}

So far, continuous GWs have not been detected. Conventionally, these are GWs that have quasi-constant frequency and quasi-infinite duration, or at least for much longer than any given observation period. As for what gives rise to these signals, there are many possibilities, but one source commonly discussed is from spinning isolated neutron stars. On neutron stars, there are several mechanisms that produce GWs including long-lasting global deformations (``mountains''), precession, and r-modes. All these mechanisms result in either a mass or current multipole moment, which ultimately results in GW emission~\cite{thorne1980}. Further information regarding these models and how we detect these signals can be found in one of the many recent reviews~\cite{sieniawskaBejger2019,haskellSchwenzer2021,piccinni2022,riles2023,wette2023,haskellBejger2023}.

In addition to quasi-infinite continuous GWs, there is the idea of \textit{transient} continuous GWs~\cite{prixetal2011}. Trivially, these refer to signals with finite durations, typically between hours to months, but retain the attribute of having quasi-constant frequency. One astrophysical phenomenon that could be related to these types of signals is pulsar glitches, where a neutron star is observed to rapidly increase its spin which is sometimes followed by a recovery that takes several days to months. GW models include the free energy released during superfluid vortex unpinning \cite{prixetal2011}, Ekman flow \cite{vanEysdenMelatos2008}, transient mountains \cite{yimJones2020}, and recently, trapped ejecta \cite{yimetal2024}. Further information can be found in the recent review by Haskell \& Jones~\cite{haskellJones2024}.

\paragraph{\textbf{Lorenzo Speri} -- Gravitational self-force: The two-body problem in the small mass ratio limit} \mbox{}

GW modeling is a complex field involving solving Einstein's field equations to predict the relationship between source parameters and waveform signals. The GW self-force program focuses on solving the two-body problem under the small mass ratio approximation, typically seen in scenarios like inspirals of small objects around SMBHs at galaxy centers, known as EMRIs, a prime target for future space-based GW detectors~\cite{barack_2019,babak_2017}. The small mass ratio approximation is particularly useful for studying asymmetric binaries, which are challenging to simulate using Numerical Relativity (NR) codes~\cite{wittek_2024}. Filling this parameter space aids waveform modeling and other approaches to solving Einstein's field equations. Asymmetric binaries, characterized by longer inspirals, offer precise parameter measurements due to accumulated cycles.
 
The small mass ratio approximation involves three aspects: expansion of Einstein's field equations to obtain the metric, describing the secondary object's evolution through geodesic motion affected by self-force, and simplifying the evolution description using the two-time scale approximation~\cite{hinderer_2008}. This approximation divides the problem into orbital and radiation-reaction timescales, streamlining the solution process. The self-force program has successfully produced waveforms up to second order in mass ratio for Schwarzschild BHs, matching NR simulations~\cite{warburton_2021,wardell_2023,wittek_2023,wittek_2024}. Generating EMRI waveforms poses challenges due to extended signal duration and complex harmonic content~\cite{burke_2023}. The duration of the inspiral phase is inversely proportional to the mass ratio, leading to computationally expensive waveform generation. Rich harmonic content further increases computational costs. Despite challenges, the FastEMRIWaveform package, utilizing GPUs, has expedited waveform computations~\cite{katz_2021,speri_2023}.
 
Modeling and extracting EMRI signals from LISA data streams will provide precise binary system measurements, enabling precise tests of GR and enhancing our understanding of various astrophysical phenomena~\cite{barack_2007,yunes_2011,berry_2019,maselli_2020,speri_2023b}. Additionally, EMRIs offer avenues for constraining cosmological parameters and measuring phase calibration errors~\cite{laghi_2021,savalle_2022}. Moving forward, expanding the parameter space from simpler to more general spacetimes, and improving waveform implementations are crucial. From a data analysis perspective, exploring EMRIs remains a vital pursuit.

\paragraph{\textbf{Torben Frost} -- Gravitational lensing of GWs: Geometric optics and weak field wave optics} \mbox{}

The speaker asked to be exempted from the summary paper.

\subsection{Detection}

\paragraph{\textbf{Ziming Wang} -- Fisher-matrix method in GW parameter estimation} \mbox{}

In this talk, I introduce briefly the concept of parameter estimation and some basic knowledge about its application to GWs. The main content is the Fisher matrix (FM), a widely used tool in GW parameter estimation, which can estimate the error of the parameters very quickly. In fact, according to different interpretations of probability, there are two types of FMs. The frequentist FM $F^F$ originates from the ensemble average of the log-likelihood, $\langle\mathcal{L}\rangle$, where the log-likelihood is defined as $\mathcal{L} := -\log P(\text{Data}|\text{Parameter})$. It is also worth noting that the inverse of $F^F$ is the Cram{\'e}r-Rao lower bound for the variance of unbiased estimators. The Bayesian FM $F^B$, on the other hand, depends on the specific noise realization -- $F^B$ is defined as the Hessian of $\mathcal{L}$ at the max-likelihood point. We show a summary of the differences between the two FMs in Table~\ref{tab:fm}. Though defined differently, the two FMs are closely related and coincide in many cases: linear models, large data sets, and high SNR limits. Moreover, the frequentist FM is equal to the Bayesian FM in the zero-noise realization scenario.

\begin{table}\centering
    \begin{tabular}{c c c}
         \hline \\[-2.2ex] \hline \\[-1.7ex]
          & $F^F_{\alpha\beta}$ & $F^B_{\alpha\beta}$ \\[1ex] \hline \\[-1.7ex]
         Definition & $\langle\mathcal{L},_{\alpha\beta}\rangle$ & $\mathcal{L},_{\alpha\beta}|_\text{MLE}$ \\
         $\Tilde{\Theta}$ known? & Yes, fixed beforehand & No, needs estimation \\
         $\Hat{\Theta}$ known? & No, depends on noise & Yes, determined by data \\
         Average? & Yes & No \\
         Random? & No & Yes \\
         Application & Forecasts & Parameter estimation with real data \\[1ex] \hline \\[-2.2ex] \hline
    \end{tabular}
    \caption{Comparison of the frequentist and Bayesian Fisher matrices. The subscripts $\alpha$ and $\beta$ denote the parameter indices. The symbol $\Tilde{\Theta}$ denotes the true parameter while the symbol $\Hat{\Theta}$ denotes the maximum likelihood estimated parameter.}
    \label{tab:fm}
\end{table}

\paragraph{\textbf{Stefano Rinaldi} -- From interferometers to populations: An overview of GW data analysis} \mbox{}

GWs are detected making use of so-called \emph{gravitational interferometers}, giant versions of the tabletop experiment conceived by Michelson and Morley in 1887~\cite{ligodetector:2015,virgodetector:2015,kagradetector:2013}. These apparatuses are designed to detect the relative variation in length of the two interferometer arms induced by the passage of a GW via the interference pattern of two laser beams that traveled through the two orthogonal arms. Therefore, the raw data produced by gravitational interferometers consist of a time series of light intensity readings obtained with a photodiode. From this time series, leveraging on the knowledge of the properties of both the detector noise and the expected signals, it is possible to detect the presence of a GW signal. Moreover, once a signal has been detected, it is possible to infer the properties of the compact binary system (e.g., the masses of the two BHs and their distance) using a set of techniques that revolve around the stochastic exploration of the parameter space. This process goes under the name of \emph{parameter estimation} (see Ref.~\cite{christensen:2022} for a review). The product of this task is, for each confidently detected GW, a set of posterior samples representing the posterior probability density for the parameters of the merged binary system.

We can broadly categorize the scientific results stemming from GW into two categories:

\begin{itemize}
    \item Individual events: these results are obtained by studying the properties of a single event alone. Among others, we mention the tests of GR~\cite{tgr-lvk:2021,gennari:2023,maggio:2023}, measurement of the Hubble constant with the standard siren method~\cite{gw170817:2017,morton_2023}, and gravitational lensing\footnote{At least a pair of lensed candidates are expected to be detected during the currently ongoing Fourth Observing Run~\cite{xu:2022}.};
    \item Population studies: combining different events opens up for the galaxy catalog approach~\cite{gray:2020} and spectral sirens method~\cite{mastrogiovanni:2021} for the inference of cosmological parameters as well as the investigation of the astrophysical properties of compact objects~\cite{astrodist:2023,rinaldi:2023}.
\end{itemize}

Population studies are among the most powerful tools we have to discover new physics: whereas individual events can only (with some exceptions) rule out models, in order to understand the processes at work we have to look at the whole available population of observed objects. In practice, this means modeling the expected population distribution and then inferring the parameters of such distribution in a Bayesian framework under the assumption that all the observed events come from the same distribution. This (only apparently) simple approach can be complicated to account for more details (such as multivariate distribution to include more observable at the same time and the inclusion of selection effects) or more flexibility for the models (like with non-parametric methods).

\paragraph{\textbf{Xue-Ting Zhang} -- Long-lived BHBs GW data analysis with machine learning} \mbox{}

\begin{figure}\centering
\includegraphics[width=0.95\textwidth]{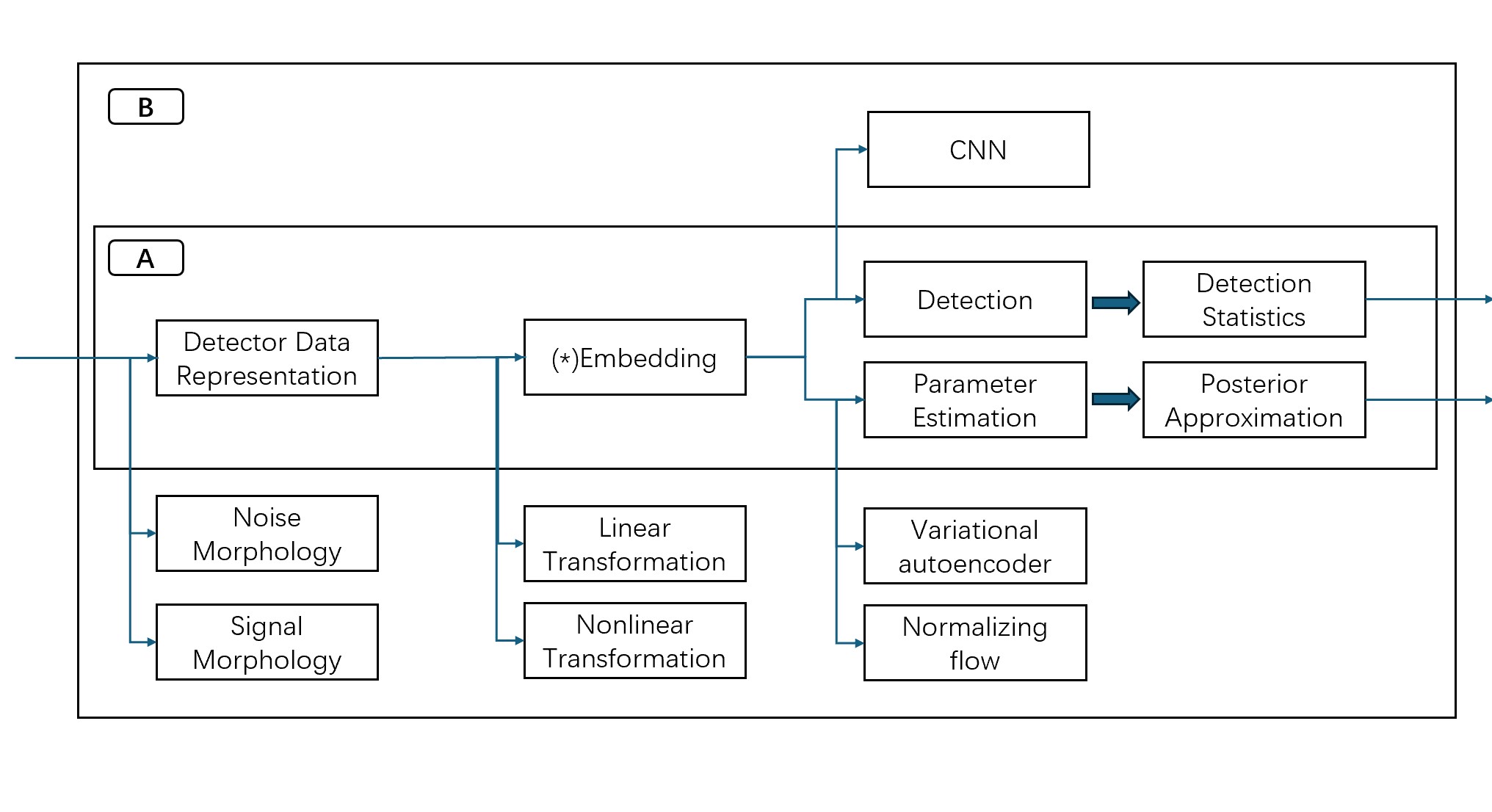}
    \caption{
        GW data analysis process involves various stages, commencing with block A which is divided into three steps critical to construct a ML pipeline. These steps include proper data representation, contemplation of data embedding, and training either a detection or parameter estimation model. For a pipeline to be effective, it must utilize these stages fully, such as proper data representation that distinguishes features between noise-only and signal-plus-noise samples, regulated by the respective noise and signal morphologies. Embedding procedures could be instrumental in deriving key insights from these preliminary features considering the complexity of the initial data for ML models. Embedding may facilitate simplified learning and mentoring ML models utilizing linear or non-linear transformation strategies. The objective of data analysis is to deploy these ML models for detection and parameter estimation tasks, creating the potential to formulate an independent and efficient pipeline. This pipeline could then serve as a valid cross-pipeline in the future, enhancing the overall analysis process.
    }\label{fig:mainstem}
\end{figure}

Machine learning (ML) has been prevalent in GW astronomy since 2017~\cite{benedetto_2023,cuoco_2021}, encompassing glitch classification, waveform modeling, GW signal detection, and parameter estimation. Here we focus on GW signal detection and parameter estimation. Taking into consideration the goals of data analysis, certain steps may require attention during this process and iterative training of suitable models, as displayed in Fig.~\ref{fig:mainstem}.

Signal detection involves classifying detector data based on the probability of data when the signal is present and absent. Convolutional neural networks (CNNs) demonstrate excellent performance in processing various GW signals, such as binary BHs~\cite{gabbard_2018,gebhard_2019,george_2018a,george_2018b,huerta_2021,schafer_2023}, binary neutron stars (BNSs)~\cite{miller_2019,krastev_2020,krastev_2021}, and EMRIs~\cite{zhang_2022,yun_2023}. These convolution kernels capture different features from the signals and score the weighted features to classify signals versus noise-only samples. Due to the complexity of the signal and non-stationary noise morphology, CNN architectures can be fine-tuned or adapted from architectures that work well in other fields, such as ResNet~\cite{nousi_2023} and LSTM~\cite{iess_2023}.

In parameter estimation, the objective is to derive the posterior of the GW parameters. Likelihood-free methods using ML can approximate these posteriors through techniques such as variational autoencoders~\cite{gabbard_2018} and normalizing flows~\cite{dax_2021,dax_2023,langendorff_2023}. Different ML architectures or algorithms have different approaches and exhibit varying potentials when aiming to construct a target distribution resembling the posterior. For instance, normalizing flows map a base distribution to a target posterior of GW parameters using an invertible mapping which is facilitated by a traceable Jacobian Matrix~\cite{dinh_2016}.

With numerous GW signals anticipated to be detected in the near future, there will be a significant occurrence of false signals attributed to noise, thereby intensifying the pressure on the data analysis pipelines. ML techniques can be effectively applied in big data and complex data environments/conditions with the assistance of GPUs and other software tools, including gradient descent algorithms to alleviate this pressure.

\section{Poster session}\label{sec:post}

A total of eight posters were presented at the meeting. However, for different reasons not all authors could provide a summary of their poster and we show here only the summary of five posters.

\paragraph{\textbf{Qian Hu} -- Rapid pre-merger localization of BNSs in third generation GW detectors} \mbox{}

\begin{figure}\centering
\includegraphics[width=0.58\textwidth]{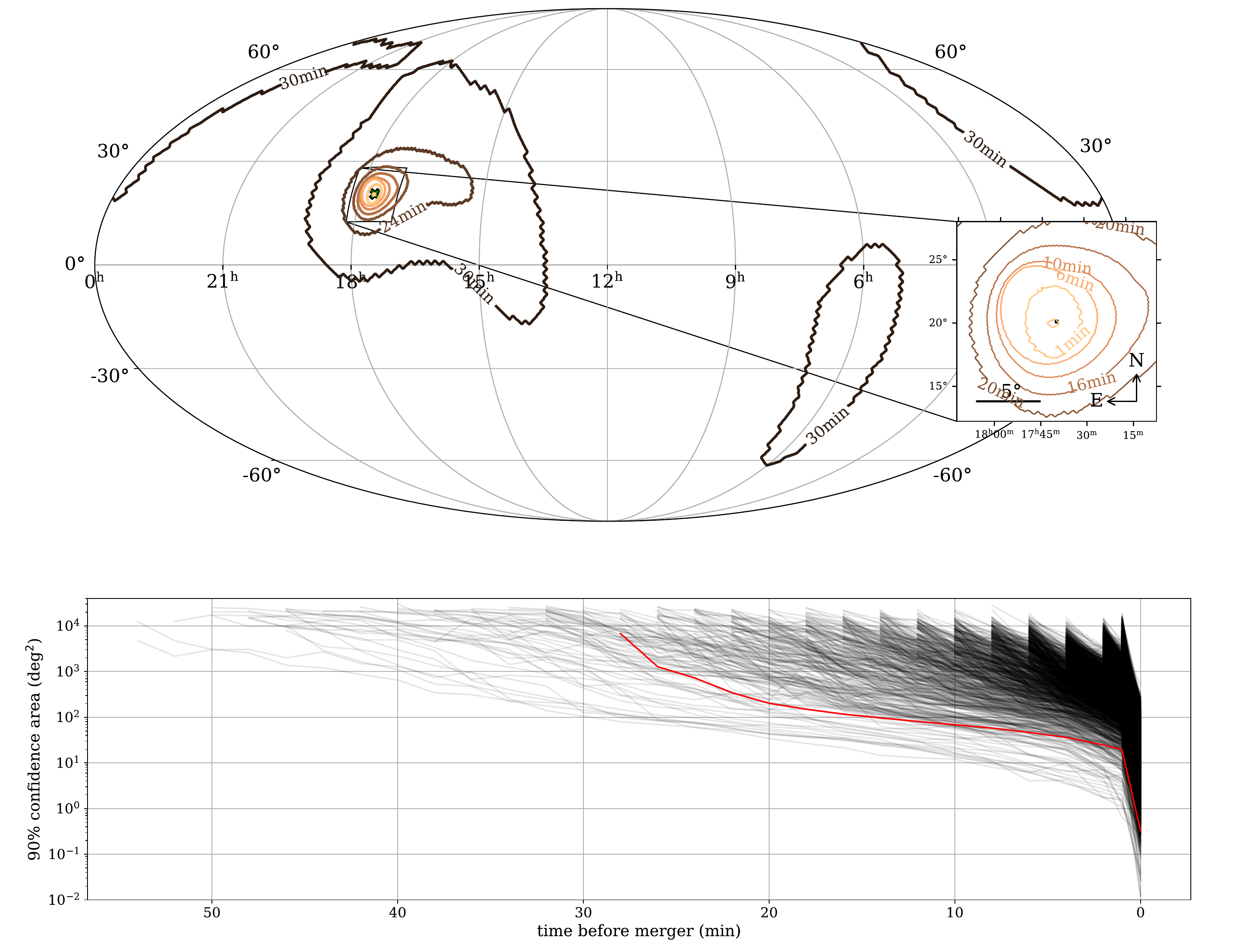}
    \caption{
        Sky map evolution for the detection accuracy of a BNS. The upper panel shows an example sky map for a $1.4+1.4\,{\rm M_\odot}$ BNS at 1000 Mpc detected 30 minutes before the merger with a network SNR of 12. The SNR increases to 17 at 20 minutes before the merger, 31 at 10 minutes, 95 at 1 minute, and 130 after the merger. We show the 90\,\% localization contours at different negative latencies. The injection sky location is marked with a cross. The lower panel shows the evolution of the 90\,\% confidence localization areas of early-warning events in our simulation. The example in the upper panel is represented by the red line. Figure from Ref.~\cite{hu_2023}.
    }\label{fig:skymap}
\end{figure}

Premerger localization of BNSs is one of the most important scientific goals for the third generation (3G) GW detectors. It will enable the EM observation of the whole process of BNS coalescence, especially for the premerger and merger phases, which have not been observed yet, opening a window for a deeper understanding of compact objects. To reach this goal, we describe a novel combination of multiband matched filtering and semi-analytical localization algorithms to achieve early-warning localization of long BNS signals in 3G detectors~\cite{hu_2023}. Using our method we are able to efficiently simulate one month of observations with a three-detector 3G network, and show that it is possible to provide accurate sky localizations more than 30 minutes before the merger. Our simulation shows that there could be $\sim10$ ($\sim1000$) BNS events localized within $100\,{\rm deg^2}$, 20 (6) minutes before the merger, per month of observation. See Fig.~\ref{fig:skymap} for more details on the evolution of sky localization.

\paragraph{\textbf{Miaoxin Liu} -- A Latin-hypercube sampling method for high-diemnsional mulitomodal distributions} \mbox{}

\begin{figure}\centering
\includegraphics[width=0.58\textwidth]{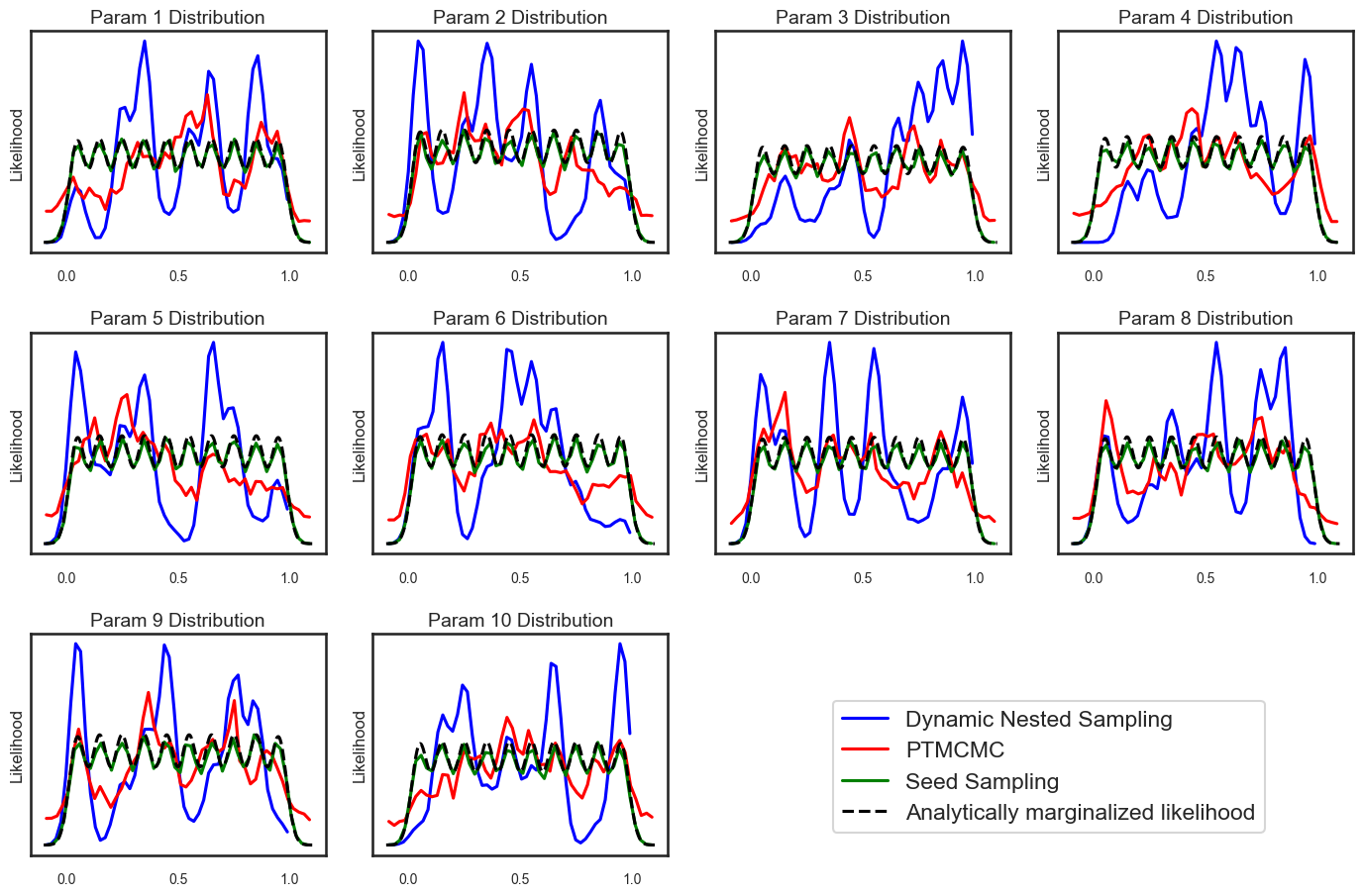}
    \caption{
        Comparison of the 1D marginalized distribution of a 10D Gaussian mixture distribution sampled by Dynesty, PTMCMC, and our method. The three methods spend 7166384, 535207, 227644 function calls, respectively. Among them, the effective sample number 108472 is the same for the three methods. The Seed Sampling result shows excellent consistency with the analytical answer compared to other methods.
    }\label{fig:10d_modes}
\end{figure}

In the rapidly expanding field of GW astronomy, efficiently analyzing high-dimensional multimodal distributions has emerged as a critical challenge, especially with the complexity introduced by sources like EMRIs and galactic binaries. To address this problem, we propose a novel sampling method named ``Seed Sampling'' that significantly enhances the efficiency and efficacy of GW source analysis. By integrating `Latin-Hypercube Sampling' with an adaptive weighting mechanism, this approach offers superior initial space exploration and focused exploration of high-likelihood regions, demonstrating clear advantages over traditional methods in terms of efficiency, adaptability, and sample utilization in multi-modal cases. Fig.~\ref{fig:10d_modes} shows the preliminary result for a 10D Gaussian mixture distribution.

\paragraph{\textbf{Xiangyu Lyu} -- Parameter estimation of stellar-mass BHBs under the network of TianQin and LISA} \mbox{}

For space-borne GW detectors, like TianQin and LISA, stellar-mass BHB system is a type of observable source. In our study, we first provided TianQin's full-frequency response function in time-delay-interferometry (TDI), selected two stellar-mass BHB systems from GWTC-3, and adjusted key parameters to meet the detection SNR threshold~\cite{lyu_2023}. Leveraging this foundation, we generated mock observation data for two years of stellar-mass BHB GW observation by TianQin and LISA. we employed a Bayesian inference framework to assess the accuracy of parameter estimation. Our findings reveal that joint observations (TianQin+LISA) generally enhance the accuracy of parameter estimation by nearly a factor of two, compared to observations made by TianQin or LISA alone. Notably, when dealing with stellar-mass BHB systems possessing non-zero spin, accurate modeling of spin effects becomes imperative in the parameter estimation process. Overlooking these spin effects can introduce significant biases in the estimation of mass parameters, amounting to a $3-\sigma$ deviation, as well as in the determination of sky location, leading to a $1-\sigma$ discrepancy. Therefore, proper consideration of spin effects is vital for obtaining accurate and reliable parameter estimation results.

\begin{figure}\centering
\includegraphics[width=0.58\textwidth]{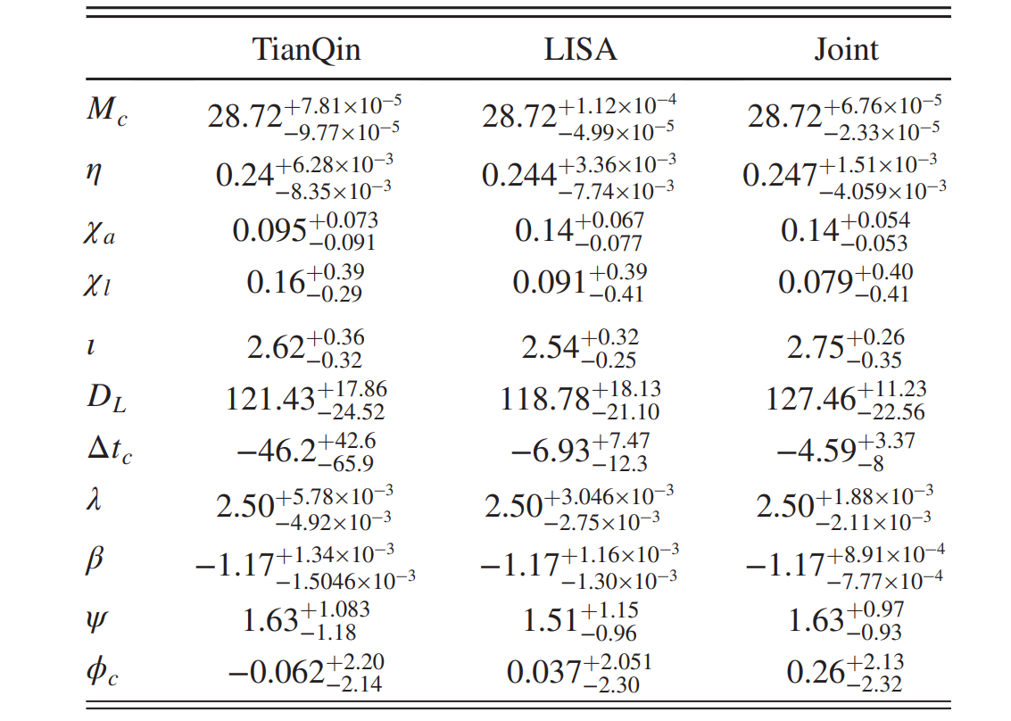}
    \caption{
        Estimated parameters for a GW150914-like source detected by TianQin alone, LISA alone, and joint detection. The parameters considered are the chirp mass $M_c$, the symmetric mass ratio $\eta$, symmetric and antisymmetric combinations of the spins $\chi_a$ and $\chi_l$, the inclination of the source $\iota$, the source's luminosity distance $D_L$, the coalescence time $t_c$, the ecliptic longitude and ecliptic latitude $(\lambda,\beta)$ in the solar-system barycenter system, the polarization angle $\psi$, and the coalescence phase $\phi_c$. 
    }\label{fig:gw150914}
\end{figure}

\paragraph{\textbf{Zheng Wu} -- Searching GW bursts with space-borne detectors} \mbox{}

\begin{figure}\centering
\includegraphics[width=0.58\textwidth]{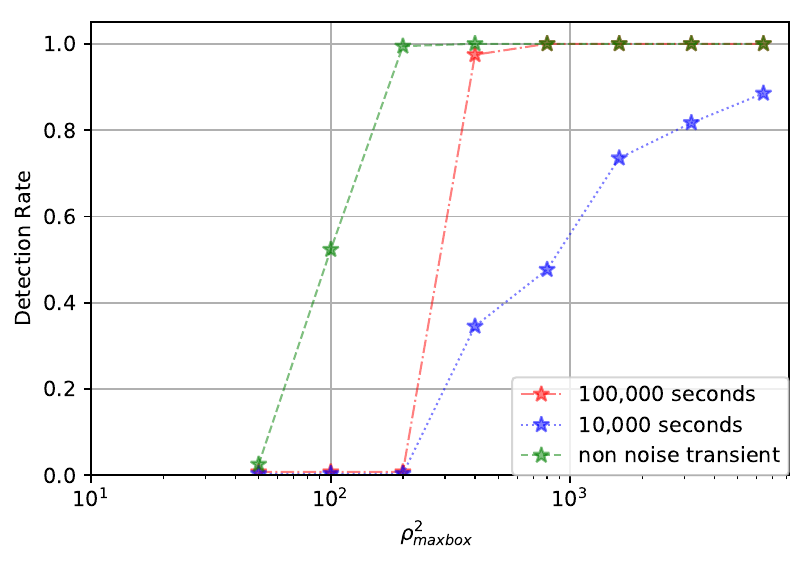}
    \caption{
        The detection rate as a function of $\rho_\text{maxbox}$ for three different noise environments: no noise transients (green), the average interval between noise transients is 100,000 seconds (orange), and the average interval between noise transients is 10,000 seconds (blue). The detection rate is 99.4\,\% for signals with $\rho_\text{maxbox} = 14.1$ in the absence of noise transients, it is 97.4\,\% for an average interval between noise transients of 100,000 seconds for signals with $\rho_\text{maxbox} = 20.0$, and the detection rate drops to 88.5\,\% for an average interval of 10,000 seconds for signals with $\rho_\text{maxbox} = 80.0$. Figure from Ref.~\cite{wu_2023}.
    }\label{fig:performance}
\end{figure}

Various mechanisms can produce short outbursts of GWs, whose actual waveform can be hard to model. In order to identify such GW bursts and not misclassify them as noise transients, we proposed a proof-of-principle power excess method that comprehensively utilized two signal channels and the signal-insensitive channel to veto noise transients~\cite{wu_2023}. The detection rate as a function of the maximum SNR box components of the burst signals $\rho_\text{maxbox}$ for different average intervals between the noise transients and no noise transients is shown in Fig.~\ref{fig:performance}.

\paragraph{\textbf{Cong Zhou} -- Probing orbits of stellar-mass objects deep in galactic nuclei with QPEs} \mbox{}

\begin{figure}\centering
\includegraphics[width=0.58\textwidth]{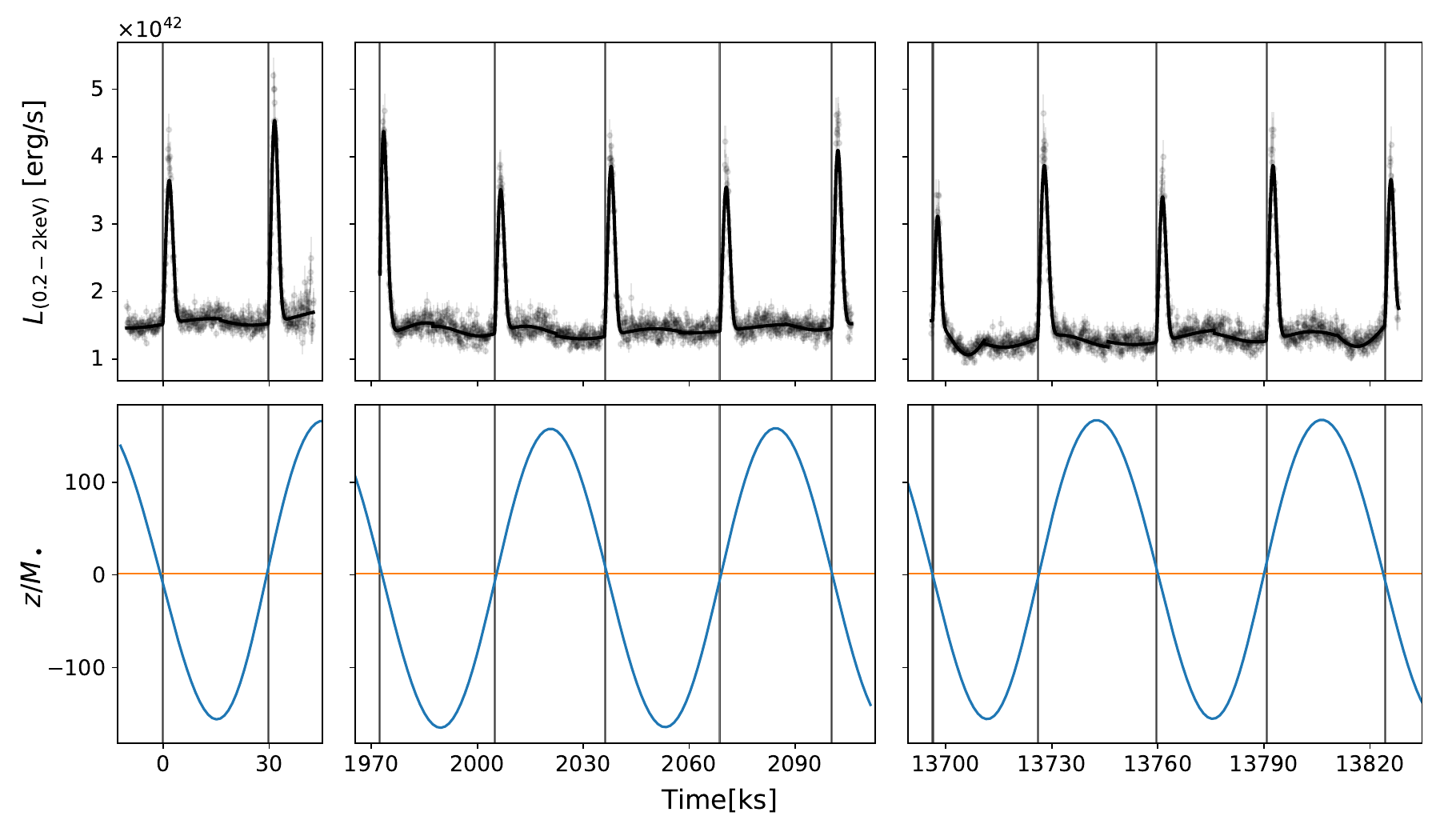}
    \caption{
        The top panel shows the light curve data along with the best fit of the emission model, where the vertical lines are the starting times of the QPEs. The bottom panel shows the distance to the equator plane of the central SMBH $z(t)$ for the best-fit orbit, where the horizontal lines denote the disk surface z=H. Figure from Ref.~\cite{zhou_2024}.
    }\label{fig:qpe}
\end{figure}

QPEs are intense repeating soft X-ray bursts with recurrence times of about a few to ten hours from nearby galactic nuclei. The origin of QPEs is still unclear. In our work, we investigated the EMRI plus accretion disk model, where the disk is formed from a previous tidal disruption event (TDE)~\cite{zhou_2024}. In this EMRI+TDE disk model, the QPEs are the result of collisions between a TDE disk and a stellar-mass object (a stellar-mass BHB or a main sequence star) orbiting around a SMBH in galactic nuclei. If this interpretation is correct, QPEs will be invaluable in probing the orbits of stellar-mass objects in the vicinity of SMBHs, and further inferring the formation of EMRIs which are one of the primary targets of spaceborne GW missions. Taking GSN 069 as an example, we find the EMRI therein to be of low eccentricity ($e < 0.1$ at $3-\sigma$ confidence level) and to have a semi-major axis of $O(10^2)$ gravitational radii from the central SMBH, which is consistent with the prediction of the wet EMRI formation channel, while incompatible with alternatives. In Fig.~\ref{fig:qpe} we show the light curve data of GSN 069 and the best-fit orbit for an EMRI.

\section{Discussion sessions}\label{sec:dis}

The discussion sessions were led by questions handed in by the participants before the meeting. The questions can be found in \ref{app:que}.

\subsection{Astrophysics}

During the discussion session on astrophysics three topics attracted major attention. The first topic was the importance of EM counterparts of GW detections. EM+GW detections will allow us to measure cosmological parameters using bright standard sirens~\cite{schutz_1986,chen_2022,morton_2023}. Although dark standard sirens are also discussed in the literature~\cite{ligo_2019,ligo_2021a}, the general consensus was that for dark standard sirens risks of unnoticed bias are much more significant. Although measurements with dark standard sirens are an important contribution to cosmology with GWs, pushing for more detections of EM counterparts is necessary. Moreover, other gains from EM detection are also expected. Tests of GR measuring the difference in the travel times of light and GWs were emphasized as one of the most reliable tests due to their relative simplicity and the good accuracy we have for measuring arrival times~\cite{wei_2017,shoemaker_2018}. Furthermore, it was highlighted that the detection of the EM counterpart can greatly improve the sky localization of the source which allows a better detection accuracy of all parameters~\cite{morton_2023}. In some sense, it was concluded that multi-messenger astronomy should not be just a ``byproduct'' of GW detection but one major focus since once again it is true that ``the whole is greater than the sum of its parts''.

The second topic that caught major attention was the issue that the models used in GW detection sometimes do not reflect the current understanding we have of astronomy and astrophysics. An example discussed is that the LIGO-Virgo-KAGRA (LVK) Collaboration uses as a general assumption that stellar-mass BHBs tend to have a mass ratio of one although astrophysical models and GW detections point toward unequal mass binaries~\cite{ligo_2021b,ligo_2023,rinaldi_2023}. In particular, the concern was raised that the number of sources with high mass ratios might be even greater but we are simply missing them due to a lack of appropriate waveform models~\cite{varma_2019,jan_2023,van-de-meent_2023}. Although it is clear that a lack of waveform models is due to the complexity of modeling sources in the intermediate mass regime, it was concluded that developing this kind of waveform should be prioritized. Another point discussed was that population models derived from LVK do not always include our understanding of stellar evolution. For example, the merger rate of BHBs at different cosmological redshift is often assumed to follow the star formation rate but a time difference should be expected. In particular when dealing with BHs of different masses because the lifetime of stars differs a lot depending on their initial mass~\cite{lamers_2017}. Furthermore, environmental effects are often neglected when interpreting GW detection. The general assumption tends to be that BHBs evolve as isolated systems but it is very likely for them to form in dense systems. In particular, the formation of BHBs in AGN disks has been discussed a lot recently which means the evolution of these sources is greatly impacted by the interaction with gas and a tertiary~\cite{ford_2022,vaccaro_2023}. Although the presence of gas should not greatly affect the detected signal it plays a major role in the evolution of the binary and thus in the merger rate. Other effects like redshifts induced due to the orbital motion of the binary around the SMBH or its gravitational pulling, however, can affect the detected signal as well as the information inferred from the signal, including merger rates since the apparent distance of the source is also affected~\cite{chen_2019,torres-orjuela_2020,torres-orjuela_2022}.

The last topic that was discussed with much interest from the audience was the high number of sources that we expect for future detectors and how they will affect detection~\cite{littenberg_2023,strub_2024}. Although the extraction of a high number of sources is mainly considered a task or issue for detection and data analysis, it was pointed out that astrophysics could help solve this problem. On the one hand, there are still big uncertainties on the number of sources we can expect in future detectors; in particular, for space-based detectors~\cite{lisa_2022,torres-orjuela_2023}. If we could refine our estimation of the number of sources, the schemes being developed for the search of multiple sources could be adapted better to the number of sources expected. Moreover, different kinds of sources require different treatments when dealing with their signals and hence knowing the ratio between the different sources would also help specify the properties of search algorithms~\cite{keitel_2021,ye_2023,wu_2023}. Another issue is that for many sources there is great uncertainty about their parameters which can also greatly affect their detection. As an example, we do not know very well if we can expect to have mainly highly eccentric or mainly circular EMRIs but the eccentricity greatly affects the number of modes a source has and therefore might require very different approaches~\cite{pan_2021a,pan_2021b}. It was further pointed out that some sources, that are not necessarily the most prominently discussed, might have a very important (maybe even dominant) contribution to the signal detected~\cite{amaro-seoane_2019,vazquez-aceves_2022,amaro-seoane_2024}. These sources should be studied in more detail to make sure the overall performance of GW detection can be improved.

\subsection{GW theory}

One of the main topics discussed during the theory discussion session is how to test GR with the help of accurate detection of GWs. There are many steps to take before reaching a conclusion, and there are many potential complications in determining whether there are deviations from GR. However, the first requirement is a signal with a high enough SNR and an accurate waveform with no degeneracies. In that case, if the deviation from GR is higher than the error associated with the detection, then a potential deviation could be detected.  

The most accurate description of the merger and ringdown relies on NR simulations; it was mentioned that the NR community is aware of what needs to be included to perform more accurate searches that will allow us to test GR, for example, one of the main points discussed is the fact that the parameter space is not entirely covered and the necessity of developing models which can include high eccentricities, different spin orientations, and higher mass ratios. The current constraints in the parameter space could be preventing us from finding more complex systems that might give us a better insight into the merger and ringdown of GW sources; for this reason, different kinds of search methods can look for coherent signals without the need for previous modeling. These types of searches can complement the template-matching searches. 

Another topic that was widely discussed was the information that could be obtained from the quasi-normal modes (QNMs), which represent the characteristic vibrations of BHs after the merger. The information about the mass, spin, and mass ratio is encoded in the frequency and damping time, but the details of GR on the strong regime might be found in the QNMs. These modes decay over time, and their identification is not easy. Furthermore, it is still unclear at which point in the ringdown the search for QNMs must start, increasing the difficulty of its detection. If these QNMs are detected, tests of GR on the strong regime can be performed, and in the case of neutron stars, they will help in the search for the equation of state. In summary, our capacity to test GR and alternative theories relies on the accuracy of our detections. Nevertheless, from current observations, there are no indications that lead us to think of the existence of significant effects and/or deviations from GR. 
 
Another briefly discussed point is the use of GPUs in NR. The computational demands of NR can be very high depending on the system's parameters. In general, there are some constraints regarding, for example, the mass ratio and eccentricity of the systems that can be solved. In NR, the space-time is divided into grids in which field equations are solved at each timestep. GPUs are very efficient at computing simple repetitive tasks, which can benefit the iterative equation-solving process needed at each grid segment. However, GPUs typically have less memory than CPUs, so it is crucial to consider the limits of the GPU memory while maintaining its performance. Another issue is that the GPUs are suitable for parallel computation; however, in NR, frequent communication between the CPU and GPU is necessary to ensure good performance, so implementing GPUs is a complex task. Nevertheless, many groups are already working on how to exploit the GPU processing power to accelerate NR simulations.  

The detection of dark matter candidates and dark energy was also addressed in the discussion. Currently, the European Pulsar Timing Array (EPTA) provides some constraints on dark matter candidates based on the kind of potential that the candidates would create and the effect on the arrival of the radio pulses. The presence of dark matter in binary systems could induce small deviations from GR. However, the question of how strong these deviations are is still open, and it will likely be difficult to detect dark matter only with GWs because its signal could degenerate with other effects. 

Also, GW cosmology might help to place some constraints on the universe's expansion but not so much on the nature of dark energy. It was also discussed if there is any reason to think that a GW generated at high redshift would differ from a GW generated in the nearby universe, implying that the GWs could reflect the universe's evolution. In this sense, understanding the GW background could give us some hints about it. Regardless, the current precision achieved with GW cosmology is not high enough to rule out other cosmological constraints, so the models must be thoroughly tested before claiming any deviation. 

During the discussion, there was a shift towards considering more speculative questions; the possibility of having systems that produce non-GR waveforms was addressed, and the conclusion was that we would not be able to find such systems with our current methods. The existence of such systems is not expected as currently there is no indication of non-GR systems; however, this reasoning leads to the question of whether there are effects that could lead to significant deviations from GR that might not only be observable, but that could cause a complete misinterpretation of the signal or even prevent its identification, for example in extreme environments in which quantum effects can be equally dominant compared to GR. Other speculative topics, such as the possibility of GWs collapsing into a BH or the effect of a highly curved space-time on the propagation of a GW, were also discussed; these topics remain a pure curiosity.

\subsection{Detection}

First, as discussed in the Astrophysics session, detecting EM counterparts would come with many benefits. However, to increase the number of possible EM counterparts detected, it is necessary to obtain reliable forecasts for mergers. Two points were highlighted when discussing this goal. On the one hand, warnings should be as early as possible; in the best case before the merger. For high-frequency sources, such an early warning is very ambitious due to the short time the signals spend in the band. However, it was discussed that search algorithms using ML could greatly accelerate detection as well as the communication of early warnings. On the other hand, sky localization needs to be accurate enough to guarantee the detection of the EM counterpart with high resolution. The possibility of using trained algorithms that not only include the information from the GW signal but also information about the location of likely host galaxies was also discussed.

Then, using ML to accelerate GW data analysis was discussed. ML offers powerful tools to accelerate GW data analysis in various ways. For signal detection, ML algorithms, especially deep learning models like CNNs, can be trained to recognize the characteristic patterns of GW signals in noisy data. This can lead to faster and more reliable detection of GW events compared to traditional methods. For parameter estimation, once a GW signal is detected, ML techniques can be used to estimate the physical parameters of the source, such as the masses and spins of the merging objects. A well-trained neural network can provide rapid parameter estimation, which is crucial for timely follow-up observations with other telescopes. Once optimized for fast execution, ML models enable real-time analysis of GW data, which is particularly important for identifying and responding to transient GW events, such as neutron star mergers that may have EM counterparts observable by other telescopes. Other relevant applications for ML include noise reduction, data management, and anomaly detection. By training to distinguish between genuine GW signals and various sources of noise in the detectors, techniques like autoencoders can be used for denoising, improving the SNR, and enhancing the detectability of weak signals. ML can also help manage the vast amounts of data produced by GW detectors. For example, unsupervised learning techniques can be used for data clustering and compression, making it easier to store and process the data. Last but not least, ML can be employed to identify unusual or unexpected features in GW data, potentially leading to the discovery of new astrophysical phenomena or insights into the behavior of the detectors. In summary, the use of ML in GW data analysis has the potential to significantly improve the speed and accuracy of detecting and interpreting GW signals, opening up new possibilities for astrophysical discoveries.

The third topic discussed was what we can do now to prepare for the space-borne GW detectors. Space-borne GW detectors, like LISA, TianQin, and Taiji, will open new windows on the universe by observing GWs at low frequencies not accessible to current ground-based detectors, enabling the study of SMBH mergers, EMRIs, and possibly new unexpected sources. Preparing for those missions involves various strategic, technical, and scientific efforts. The most important aspect is enhancing data analysis techniques, which include generating large amounts of accurate templates, developing ML algorithms, and improving computational infrastructure. To generate accurate templates, refining theoretical models of GW sources to increase the accuracy of signal waveforms is crucial. Then extensive simulations of GW signals expected to be detected by the space-borne missions need to be performed to make sure that they can be extracted in data analysis. Development of ML and deep learning algorithms for signal detection, noise reduction, and parameter estimation specific to the low-frequency GW signals space-borne detectors is also necessary. In the meantime, we also need to consider building and enhancing computational infrastructure capable of handling the vast amounts of data that space-borne detectors will generate. Other essential aspects to consider in preparing for the space-borne GW detectors include strengthening multi-messenger astronomy, fostering international collaboration, and developing technology and engineering solutions to space environmental challenges. Multi-messenger astronomy requires building networks between GW observatories and other astronomical facilities (e.g., optical, radio, and X-ray telescopes) to follow up on GW events and to develop strategies for coordinated observations across different messengers to maximize scientific returns from GW detections. Since space-borne GW observatories will be major international efforts, strengthening existing partnerships and forming new ones will be key to their success. We need to promote the exchange of knowledge, data, and techniques among international research groups and institutions to foster a global community of GW researchers. In addition to the collaborations, continue developing and testing the technologies that will be used in space-borne GW detectors, such as laser interferometry, drag-free control, and precision formation flying, and devising solutions to mitigate the impact of cosmic rays, solar wind, and other space environmental factors on detector sensitivity, are also crucial for those missions to be successfully carried out. 

Finally, the detection of GWs using PTAs and how to identify their sources were discussed. PTAs aim to detect GWs by monitoring the precise time of arrival of pulses from multiple millisecond pulsars distributed across the sky. These pulsars are incredibly stable natural clocks, and GWs passing between Earth and the pulsars can cause detectable variations in the pulse arrival times. PTAs are particularly sensitive to very low-frequency GWs (in the nHz range), making them ideal for detecting waves emitted by SMBH binaries, cosmic strings, and other massive cosmological phenomena that emit GWs at these frequencies. In 2023, several PTA groups, including the EPTA in collaboration with the Indian Pulsar Timing Array (InPTA), the North American Nanohertz Observatory for GWs (NANOGrav), the Parkes Pulsar Timing Array (PPTA), and the Chinese Pulsar Timing Array (CPTA), announced a potential discovery that is likely to be a GW signal whose source has not been identified. Progress in further reducing noises and increasing the sensitivity of PTAs are crucial for identifying the nature of potential sources for GWs detected by future PTAs. International collaborations are also significant in improving detection capabilities. In the meantime, comprehensive calculations in theoretical models for GWs from the early universe need to be carried out to compare with future PTA data. As the sensitivity of PTAs improves, they will complement ground-based and space-borne GW observatories, providing a more complete picture of the universe's GW background.

\section{Summary}\label{sec:sum}

During the meeting, lively discussions on the theory and detection of GWs as well as the astrophysics governing the formation and evolution of the sources were conducted. The general consensus was that despite the great achievements that have been achieved in GW astronomy, there is still a long way to go promising interesting challenges and exciting discoveries. One major open issue is that the models used in GW detection often do not reflect our current understanding of the underlying astrophysics: e.g., the delay between the formation of massive stars and BHBs or environmental effects. Another point discussed was that template banks are not complete which can lead to the misinterpretation of sources or that we do not detect them at all. Although this is mainly a problem concerning GW detection, astrophysics and theory can greatly contribute to solving this issue by giving an understanding of what properties we could expect/should focus on for GW sources or expanding our understanding of alternative theories of gravity. Testing alternative theories and containing GR was highlighted as a major goal of GW astronomy. Here it was discussed that we require a better understanding of the astronomical environments of the sources to understand potential confusion or bias for different effects. Moreover, the development of appropriate waveform models that can handle new effects and integrate our understanding of related astrophysics remains crucial. For tests of GR as well as for cosmological measurements, the value of EM counterparts was highlighted. Multi-messenger detections were discussed as being particularly valuable for tests of GR as they allow relatively simple and reliable tests. It was further emphasized that detecting EM counterparts has many more benefits than the aforementioned ones, as they also allow constraining the properties of the source more accurately as well as studying its environment. However, the detection of EM counterparts requires a significant improvement in detecting and measuring the properties of the sources. It was discussed that all tools available should be utilized to achieve this goal, including the use of ML algorithms and similar techniques.

\ack
We thank the Kavli Institute for Astronomy and Astrophysics at Peking University, the Department of Astronomy at Tsinghua University, the Scientific Organizing Committee, the Advisory Committee, and the staff who made this workshop possible. This meeting was supported by the China Postdoctoral Science Foundation (No. 2023M741999) and the National Science Foundation of China (Grant  No. 11991053).

\appendix

\section{Questions}\label{app:que}

Here, we show a complete list of the questions submitted by the participants.

\subsection{Astrophysics}

\begin{itemize}
    \item From the point of view of astrophysical models, what is missing yet? And what, instead, can we assume as robustly grounded currently?
    \item Is it a promising research area to study the binary compact objects affected by the third body perturbation?
    \item How often do we expect BHBs around a SMBH?
    \item What are the most promising observational signatures to help us disentangle the various merger channels?
    \item Do we have an unambiguous smoking gun to claim that we observed a specific formation channel?
    \item What is the unique signature to constrain the formation channels of BHBs in LISA/Tianqin/Taiji detection?
    \item What is the significance of constraining the eccentricity of the BHB before its merger?
    \item What is the time delay between the formation of massive stars and the merger of BHBs? Any models?
    \item What causes the merger of SMBHs?
    \item Are the PTA band stochastic GWs of primordial or astrophysical nature?
    \item How can we distinguish astrophysical and cosmological sources in PTA?
    \item How can we utilize the increasing information from high-frequency detections to prepare for the space-based generation?
    \item On what tasks in GW astronomy should ML be applied?
    \item What are the most promising IMBH merger channels?
    \item What are the most interesting astrophysical applications of EMRIs?
    \item What are possible ways to form an EM counterpart to GW sources?
    \item Can multi-messenger astrophysical signals be used to examine GW waveform?
    \item How can we leverage multi-messenger astronomy to enhance our understanding of GW sources and their environments?
    \item What is the new GW source after we have found BHB, BH-NS, and BNS?
    \item The next's next: What is the prospect after 3G era?
    \item What is the new big discovery for future GW detections?
    \item What aspect of GW experiments has the highest environmental impact?
    \item How to work in large GW collaborations?
\end{itemize}

\subsection{GW theory}

\begin{itemize}
    \item How can we improve the theories from GW observations and what can we learn from the theories?
    \item What is the new big discovery for future GW detections?
    \item What relativistic effects are the most important in gravitational waveform modeling?
    \item From the perspective of relativity, what effects must be taken into consideration?
    \item To what extent can GWs aid our study of neutron star matter?
    \item What are the implications of GW astronomy for fundamental physics, such as testing general relativity or probing the nature of dark matter and dark energy?
    \item Can we use GWs from distant sources to detect the effects of quantum gravity?
    \item How can we use continuous GWs to test gravity theories?
    \item When using the current LVK network to study modified theories of gravity, what is the most important thing, long-term observation data, more loud events, or something else?
    \item What is the most hopeful candidate for modified gravity theories?
    \item Do GWs in spherical coordinates carry additional information about different degrees of freedom?
    \item How much information about GWs is contained in QNMs?
    \item How do GWs interact with each other in the case of a strongly curved background spacetime?
    \item Can overlapping GWs collapse into a BH?
    \item How do ultralight particles (such as axions) around a Kerr BH produce GWs?
    \item Can GPUs be used in NR and how?
\end{itemize}

\subsection{Detection}

\begin{itemize}
    \item What are the most pressing unsolved questions that GW detectors can address in the coming decade?
    \item What is the new big discovery for future GW detections?
    \item What are the key scientific objectives we want to address? Which aspects of waveform modeling do we need to focus on to achieve these objectives?
    \item How to figure out new physics in GW data?
    \item Now that we are able to detect around one BHB signal per day, what is there left to do for researchers in the field?
    \item How can we make it easier to observe the EM counterparts of GW events?
    \item How can we combine GW with other probes (EM, neutrino, etc) in the best way?
    \item What are potential systematic errors in multi-messenger observations (GW and EM) of SMBH binaries or stellar-mass BHBs in AGNs? How can we avoid them?
    \item What is the key question to distinguish different formation channels?
    \item How can we get a proper re-scaling of the target distribution?
    \item How to accelerate GW template generation and data analysis?
    \item How to push ML techniques into real usage of GW data analysis?
    \item How can we efficiently detect and confirm the abundance of signals that are detectable by space-borne GW detectors?
    \item What is the main challenge in LISA data processing?
    \item Compared to ground-based detectors like LIGO, what are the main challenges faced by space-based GW detectors?
    \item How to quickly search for and estimate the parameters of the stronger signals in multi-source mixing?
    \item In the frequency band of space GW detectors, do we have an efficient pipeline to estimate the parameters of BNS coalescences that last for several days?
    \item How can we compute the time-domain likelihood function more quickly?
    \item What is the progress of Taiji and TianQin?
    \item What are the key advantages and disadvantages of Tianqin vs Taiji?
    \item What synergies can we expect between LISA and China’s space-based GW detectors?
    \item When the GW arrives at the detector, how does it work from the movement of interference fringes to the final detected GW strains?
    \item What are the requirements for waveforms?
    \item How do we define accuracy for a waveform model?
    \item Assuming we have the theory behind the waveform modeling of interest, is the implementation going to be fast enough for analyzing the data?
    \item What do we get from the detection of higher-order multipoles in GW signals from EMRIs and IMRIs?
    \item The concept of ``spectral lines'' might be valuable in GW astronomy in the future. Can we define such a concept?
    \item How to extract the information of QNMs from the GW waveform?
    \item Are there alternative ways of detecting GWs besides laser interferometry?
    \item Is lunar GW detection possible and what advantages does it have compared to space-based detectors?
    \item How soon will pulsar-timing arrays detect GWs from individual sources?
    \item What are the prospects for low-latency real-time detections of GW signals for both ground-based and space-based detectors?
\end{itemize}

\section{Affiliations}

Manuel Arca Sedda: Gran Sasso Science Institute (GSSI), Viale Francesco Crispi 7, 67100, L’Aquila, Italy \\
INFN, Laboratori Nazionali del Gran Sasso, I-67100 Assergi, Italy \\
Yan-Chen Bi: CAS Key Laboratory of Theoretical Physics, Institute of Theoretical Physics, Chinese Academy of Sciences, Beijing 100190, China \\
School of Physical Sciences, University of Chinese Academy of Sciences, No. 19A Yuquan Road, Beijing 100049, China \\
Jin-Hong Chen: Department of Physics, The University of Hong Kong, Pokfulam Road, Hong Kong \\
Hong-Yu Chen: TianQin Research Center for Gravitational Physics and School of Physics and Astronomy, Sun Yat-sen University (Zhuhai Campus), Zhuhai 519082, China \\
Xian Chen: Department of Astronomy, School of Physics, Peking University, 100871 Beijing, China \\
Andrea Derdzinski: Department of Life and Physical Sciences, Fisk University, 1000 17th Avenue N., Nashville, TN 37208, USA \\
Department of Physics \& Astronomy, Vanderbilt University, 2301 Vanderbilt Place, Nashville, TN 37235, USA \\
Qian Hu: Institute for Gravitational Research, School of Physics and Astronomy, University of Glasgow, Glasgow, G12 8QQ, United Kingdom \\
Jiageng Jiao: International Centre for Theoretical Physics Asia-Pacific, University of Chinese Academy of Sciences, Beijing 100049, China \\
Matthias U. Kruckow: D{\'e}partement d'Astronomie, Universit{\'e} de Gen{\`e}ve, Chemin Pegasi 51, CH-1290 Versoix, Switzerland \\
Miaoxin Liu: Department of Physics, National University of Singapore, Singapore 117551 \\
Xiangyu Lyu: TianQin Research Center for Gravitational Physics and School of Physics and Astronomy, Sun Yat-sen University (Zhuhai Campus), Zhuhai 519082, China \\
Stefano Rinaldi: Institut f{\"u}r Theoretische Astrophysik, ZAH, Universit{\"a}t Heidelberg, Albert-Ueberle-Straße 2, Heidelberg, D-69120, Germany \\
Dipartimento di Fisica e Astronomia ``G. Galilei'', Universit{\`a} di Padova, Via Marzolo 8,
Padova, I-35131, Italy \\
Lorenzo Speri: Max Planck Institute for Gravitational Physics (Albert Einstein Institute), D-14476 Potsdam, Germany \\
Alejandro Torres-Orjuela: Department of Physics, The University of Hong Kong, Pokfulam Road, Hong Kong \\
Ver{\'o}nica V{\'a}zquez-Aceves: Kavli Institute for Astronomy and Astrophysics, Peking University, 100871 Beijing, China \\
Ziming Wang: Department of Astronomy, School of Physics, Peking University, 100871 Beijing, China \\
Rui Xu: Department of Astronomy, Tsinghua University, Beijing 100084, China \\
Yu-Mei Wu: School of Fundamental Physics and Mathematical Sciences, Hangzhou Institute for Advanced Study, UCAS, Hangzhou 310024, China \\
Zheng Wu: TianQin Research Center for Gravitational Physics and School of Physics and Astronomy, Sun Yat-sen University (Zhuhai Campus), Zhuhai 519082, China \\
Garvin Yim: Kavli Institute for Astronomy and Astrophysics, Peking University, 100871 Beijing, China \\
Xue-Ting Zhang: TianQin Research Center for Gravitational Physics and School of Physics and Astronomy, Sun Yat-sen University (Zhuhai Campus), Zhuhai 519082, China \\
Cong Zhou: CAS Key Laboratory for Research in Galaxies and Cosmology, Department of Astronomy,
University of Science and Technology of China, Hefei 230026, P. R. China \\


\section*{References}

\bibliographystyle{unsrt}
\bibliography{ref.bib}

\end{document}